# Unexpected entanglement dynamics in semidilute blends of supercoiled and ring DNA

Karthik R. Peddireddy[1], Megan Lee[1], Yuecheng Zhou[2], Serenity Adalbert[1], Sylas Anderson[1], Charles M. Schroeder[2], Rae M. Robertson-Anderson[1,*]

[1]*Department of Physics and Biophysics, University of San Diego, 5998 Alcala Park, San Diego, CA 92110, United States*

[2]*Department of Materials Science and Engineering, Beckman Institute for Advanced Science and Technology & Department of Chemical and Biomolecular Engineering, University of Illinois at Urbana-Champaign, Urbana, IL 61801, United States*

## ABSTRACT

Blends of polymers of different topologies, such as ring and supercoiled, naturally occur in biology and often exhibit emergent viscoelastic properties coveted in industry. However, due to their complexity, along with the difficulty of producing polymers of different topologies, the dynamics of topological polymer blends remains poorly understood. We address this void by using both passive and active microrheology to characterize the linear and nonlinear rheological properties of blends of relaxed circular and supercoiled DNA. We characterize the dynamics as we vary the concentration from below the overlap concentration $c^*$ to above (0.5$c^*$ to 2$c^*$). Surprisingly, despite working at the dilute-semidilute crossover, entanglement dynamics, such as shear-thinning, elastic plateaus, and multiple relaxation modes, emerge. Finally, blends exhibit an unexpected sustained elastic response to nonlinear strains not previously observed even in well-entangled linear polymer solutions.

## INTRODUCTION

DNA is a ubiquitous biopolymer that naturally exists in multiple topologies such as linear, relaxed circular (ring), and supercoiled.[1-4] Due to the unique ability to produce precise lengths and topologies on demand, DNA has been studied extensively over the past few decades as a model system to shed light on controversial polymer physics principles.[5-18] These studies – along with theoretical investigations and synthetic polymer experiments – have enabled a robust understanding of the dynamics of solutions of linear polymers in all three concentration regimes: dilute ($c<c^*$), semidilute ($c\sim c^*$) and entangled ($c>>c^*$), where $c^*$ is the concentration at which polymer coils begin to overlap, defined as $(3/4\pi)M/N_A R_G^3$ where $R_G$ is radius of gyration and $M$ is molecular weight.[17, 19-21] However, much less understood are the dynamics of solutions of polymers of different topologies, such as ring and supercoiled constructs, as well as polymer blends.[3, 17, 20, 22-26] Moreover, the limited studies on these systems have shown that polymeric blends can display unique and surprising viscoelastic properties that are not only intriguing from a physics point of view but also beneficial for the design of new multifunctional materials.[17, 20, 27-29] For example, blends of ring and linear polymers have been shown to display increased viscosity, suppressed relaxation, and hindered diffusion compared to monodisperse systems of linear chains or rings.[28, 30-34] These results suggest that interactions between topologically distinct polymers are key to emergent mechanics, and could be harnessed to produce tunable materials with a wide parameter space of function. However, the emergent properties reported thus far have only been observed at concentrations above the entanglement concentration $c_e$ which is several times larger than $c^*$.[13, 17, 27, 28]

Here, we combine passive and active microrheology to determine the linear and nonlinear rheological properties of blended solutions of ring and supercoiled DNA (Fig. 1). We show that these blends exhibit surprising signatures of classical polymer entanglements at concentrations much lower than similar monodisperse systems of linear or ring polymers. These emergent properties demonstrate that topological blends can be exploited to create robust and stiff materials with much lower concentrations than monodisperse systems. We hope our surprising results spark theoretical investigations to elucidate the interactions between topologically-distinct polymers that give rise to the emergent phenomena.

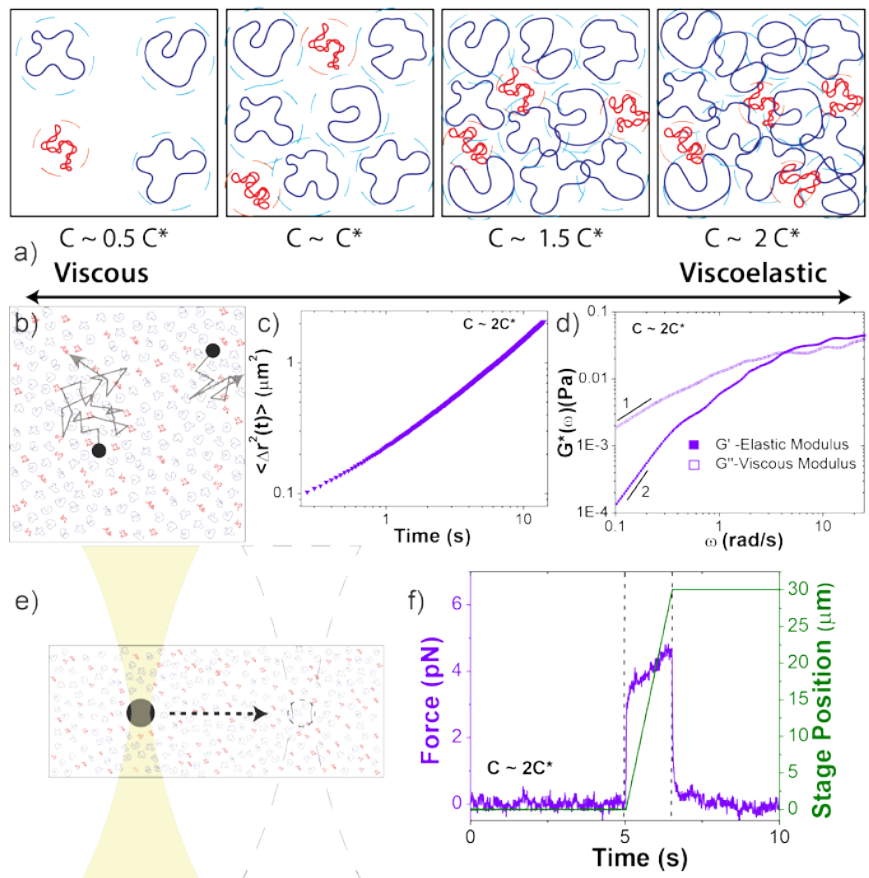

**Figure 1. Experimental approach to probe the rheological properties of blends of ring and supercoiled DNA in the dilute-semidilute crossover regime.** (a) Cartoon of blends of supercoiled (red) and ring (blue) DNA at four different concentrations that straddle the overlap concentration $c^*$. Dashed circle around each polymer coil represents its area of influence. (b-d) Passive microrheology. (b) Cartoon of 1-µm microspheres diffusing through a DNA blend. Relative sizes of DNA and beads are approximately to scale. Particle-tracking algorithms determine the frame-to-frame displacements of beads. (c) Mean-squared displacements $<\Delta r^2(t)>$ are determined from the trajectories of ~2000 beads for each blend. (d) $<\Delta r^2(t)>$ is used to determine the frequency-dependent elastic and viscous moduli, $G'(\omega)$ and $G''(\omega)$. Scaling bars indicate power-law exponents predicted for the terminal regime. (e-f) Active microrheology. (e) An optically trapped 4.5-µm bead is displaced 30 µm through each blend at speeds $v$ = 5–200 µm/s, corresponding to strain rates $\dot{\gamma} = 3v/\sqrt{2}R = 4.9 – 189 s^{-1}$ where $R$ is the bead radius. Approximate polymer sizes are increased ~$4x$ for better visibility. (f) Stage position (green) and force exerted on the trapped bead (orange) before (5 s), during (0.15-6 s), and following (9-15 s) the bead displacement (delineated by dashed lines) are recorded at 20 kHz. Data shown is for $v$ = 20 µm/s.

To frame our results, we provide a brief summary of current understanding of polymer solution dynamics.[4, 7, 17-19, 35, 36] In the dilute regime ($c \ll c^*$) the Zimm model, which accounts for hydrodynamic effects, describes dynamics.[37] Polymers in this regime are predicted to relax over the Zimm time $\tau_z = \eta_s R_G^3/k_B T$ where $\eta_s$ is solvent viscosity.[35] The storage and loss moduli, $G'(\omega)$ and $G''(\omega)$, are predicted to scale with frequency as $G'(\omega) \sim \omega^2$ and $G''(\omega) \sim \omega$ with $G'' > G'$. At higher concentrations, this scaling holds for timescales above the longest relaxation time $\tau$ (i.e. the terminal regime). As $c$ approaches $c^*$ polymers begin to overlap and the Rouse model describes dynamics. In this semidilute regime ($c \sim c^*$) solutions exhibit viscoelasticity with $G' \sim G'' \sim \omega^{1/2}$ and modest viscosity shear thinning $\eta \sim \omega^{-1/2}$.[7, 8, 38] The primary mode of stress relaxation is elastic retraction which occurs over the Rouse time $\tau_R = 6R_G^2/3\pi^2 D$.[7, 9, 35, 39, 40] Once $c$ reaches $c_e$ polymers become entangled and the reptation model describes dynamics.[19, 35, 36] The longest predicted relaxation time in this regime is the disengagement time $\tau_D = (18R_G^2/a^2)\tau_R$ where $a$ is the entanglement tube radius. For $\tau_R < t < \tau_D$, $G' > G''$ with $G'$ exhibiting a frequency-independent plateau $G_0$ while $G''$ transitions from $\omega^1$ to $\omega^{-1/4}$ scaling.[41] The crossover frequency $\omega_c$ at which $G' > G''$ provides a measure of $\tau_D$. Entangled solutions also exhibit stronger shear thinning than semidilute unentangled solutions with $\eta \sim \omega^{-(\sim 0.7-1)}$.[9, 13, 42, 43]

The dynamics of ring polymers is far more controversial due to their lack of free ends required for classical reptation theory.[17] In the semidilute regime, ring polymer solutions have been reported to have zero shear viscosities $\sim 2x$ lower than their linear counterparts with diffusion coefficients that obey Rouse scaling.[13, 17] In the nominally entangled regime, rings show no $G'$ plateau and instead exhibit scaling $G' \sim G'' \sim \omega^{0.4-0.5}$ similar to semidilute linear chains.[17, 44, 45] However, when linear polymer 'contaminants' are present, a plateau modulus is again observed as well as viscosities up to $\sim 2.5x$ larger than for linear polymers.[17] Further, $\tau_D$ for rings has been predicted and observed to be shorter than that for linear chains with $\tau_{D,R}/\tau_{D,L} = (a_R/a_L)^2(L/2p)^{-1/2}$ where $p$ is persistence length.[39, 46, 47] Finally, while some studies report terminal regime scaling for entangled rings, others show no signs of reaching the terminal regime.[44, 45] Even less is understood regarding supercoiled polymers or blends of rings and supercoils, with no rheology data or predictions to our knowledge. Importantly, due to smaller $R_G$ values,[4, 5] $c^*$ is concomitantly larger for rings and supercoils than for linear polymers of equal length (see SI).

Below we present the microrheological properties of topological DNA blends in which we fix the ratio of rings to supercoiled molecules ($R:S \approx 3:1$) and vary solution concentration from $\sim 2x$ below to $\sim 2x$ above $c^*$. We show that blends display a crossover at $\sim c^*$ to a regime with dynamics that can be described by predictions for entangled polymers, including: enhanced shear-thinning, tube disengagement, and sustained elasticity. Our results suggest that interactions between the topologically distinct polymers give rise to entanglement-like dynamics which are distinct in the linear versus nonlinear regimes.

**RESULTS AND DISCUSSION**

We first analyze trajectories of diffusing microspheres embedded in the blends to determine the dependence of the linear viscoelastic moduli on concentration. At low frequencies all blends exhibit terminal behavior with $G' \sim \omega^2$, $G'' \sim \omega^1$ and $G'' > G'$ (Fig. 2b). While this is expected for linear polymers at these modest concentrations, it contradicts recent findings for ring polymers that show no terminal regime.[45] For $c > c^*$, a crossover to $G' > G''$ is observed at frequencies of $\omega_c = 17$rad/s and $\omega_c = 4.25$rad/s for $1.5c^*$ and $2c^*$, corresponding to disengagement times $\tau_D \approx 0.4$s and $\tau_D \approx 1.5$s (Fig 2d). Surprisingly, these times are close to $\tau_D$ for comparable linear DNA systems with reported values of $\sim 0.7$s and $\sim 1.24$s.[11] In contrast, predicted values for rings are an order of magnitude smaller ($\sim 0.05$s, $\sim 0.07$s). Similarly, zero-shear viscosities for 1.5$c^*$ and 2$c^*$, determined from the low-frequency plateau in $\eta^*(\omega)$ (Fig. 2c), are markedly similar to reported values for comparable linear DNA systems,[11] while $\eta_0$ for rings is predicted to be ≥2$x$ smaller.[17]

Further, a shift in scaling of $\eta_0$ with concentration is also observed for $c>c^*$ (Fig. 2c,e). The agreement between $\eta_0$ values for $c>c^*$ and those from entangled linear DNA suggest the crossover is to an entanglement-dominated regime. Finally, all solutions exhibit shear-thinning, with scaling that increases with concentration (Fig. 2c,e) and exhibits a similar shift for $c>c^*$. Exponents for blends with $c<c^*$ are in line with the predicted Rouse scaling (0.5) and those reported for linear DNA up to $6c^*$ ($\approx c_e$).[8] Conversely, for $c>c^*$, scaling exponents match those reported for well-entangled linear DNA (~0.7–1).[42]

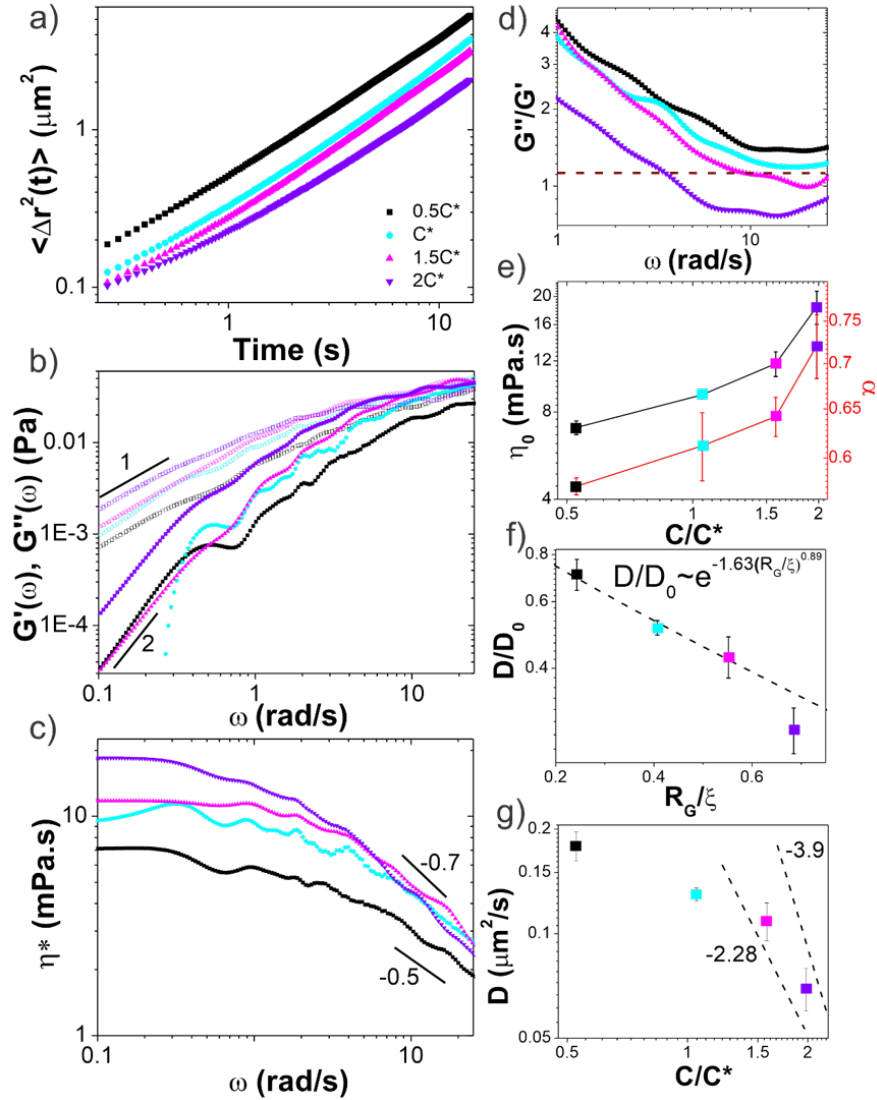

**Figure 2. Passive microrheology reveals sharp crossover in linear viscoelastic properties of ring-supercoiled blends at the overlap concentration**. (a) Mean-squared displacements $<\Delta r^2(t)>$ of microspheres diffusing through blends of $c=0.5c^*$-$2c^*$ as listed in legend. (b) Frequency-dependent elastic modulus $G'$ (closed symbols) and viscous modulus $G''$ (open symbols) determined from data shown in (a).

Scaling bars indicate power-law exponents predicted for the terminal regime ($G'\sim\omega^2$, $G''\sim\omega^1$). (c) Complex viscosity $\eta^*(\omega)$, showing varying degrees of shear thinning ($\eta\sim\omega^{-\alpha}$) with representative scaling exponents $\alpha$ shown. (d) Loss tangent ($G''/G'$) versus $\omega$ with dashed line indicating $G'=G''$. The disengagement time for each blend is determined by where the data crosses the dashed line (i.e. $\omega_c$). Note only blends with $c>c^*$ exhibit this crossover. (e) Zero shear viscosity $\eta_0$ and shear thinning exponent $\alpha$ versus $c/c^*$. (f) Diffusion coefficients $D$, determined via linear fits to $\langle\Delta r^2(t)\rangle$ (shown in (a)) and normalized by the value in buffer conditions $D_0$, plotted versus $R_G/\xi$. The dashed line corresponds to the previously reported relation $D/D_0 \sim exp(-1.63(R_G/\xi)^{0.89})$ for particles diffusing in unentangled semidilute linear polymer solutions. (g) $D$ versus $c/c^*$ with dashed lines corresponding to $D\sim(c/c^*)^{-x}$, where $x = 2.28$ and $3.9$ are the previously reported values for intermediate and large particles respectively, diffusing in entangled linear polymer solutions.

These results suggest that unexpected entanglement-like interactions occur in ring-supercoiled blends with much less coil overlap than their pure linear or ring counterparts. To corroborate this interpretation, we determine the diffusion coefficients $D$ of the particles from the mean-squared displacements (Fig 2a), and compare to predicted and empirical scalings for semidilute unentangled and entangled linear polymer solutions (Fig 2f,g).[48-51] For particle diameters $d$ comparable to the system mesh size $\xi$, and larger than or equal to $R_G$ ($d\approx\xi$, $d\geq R_G$), as in our experiments (see SI), previous studies on PEG solutions have reported the relationship $D/D_0 \sim exp(-\beta(R_G/\xi)^\delta)$, with $\beta \cong 1.63$ and $\delta \cong 0.89$ for unentangled semidilute solutions.[49] As shown in Fig 2f, our data for $c\leq c^*$ aligns with this scaling. For the entangled regime this same study reports $D\sim(c/c^*)^{-x}$ with $x = -2.28$ for $d<2a$ and $x = -3.9$, similar to predicted values for large particles,[48, 50] for $d>2a$. For linear DNA solutions with comparable length and concentration as our highest concentration blend, $a \approx 0.5$ μm, so $d \approx 2a$ in our experiments. As such, if blends were behaving similar to entangled linear polymers for $c>c^*$, as our rheology data suggests, then we should expect scaling in between these two values. The data shown in Fig 2g is indeed consistent with this picture.

To shed further light on these intriguing mechanical properties we turn to the nonlinear rheological response. To characterize the nonlinear viscoelastic response of the blends we optically drive a microsphere 30 μm through the blends at strain rates of $\dot{\gamma}=4.7–189 s^{-1}$. As shown in Figs. 3a and S3, all blends exhibit an initial elastic response in which the force increases linearly with strain followed by softening to a more viscous (i.e. strain-independent) regime. These general features are similar to those previously reported for entangled linear DNA and actin.[11, 52] However, the notable difference is the retained elasticity over the entire strain (Figs. 3a,S3). The previously reported systems all soften to a purely viscous response at large strains. This retained elasticity, which implies strong entanglements, is particularly surprising considering the modest concentrations. Another distinction between these blends and entangled linear biopolymer solutions is the lack of initial stress-stiffening (i.e. increasing slope of force with strain) before softening.[11, 52] Stiffening has been attributed to affine deformation in which polymers respond uniformly and align with the strain.[11, 53] The lack of stiffening in blends suggests that the response is non-uniform as circular polymers cannot as easily orient with the strain.

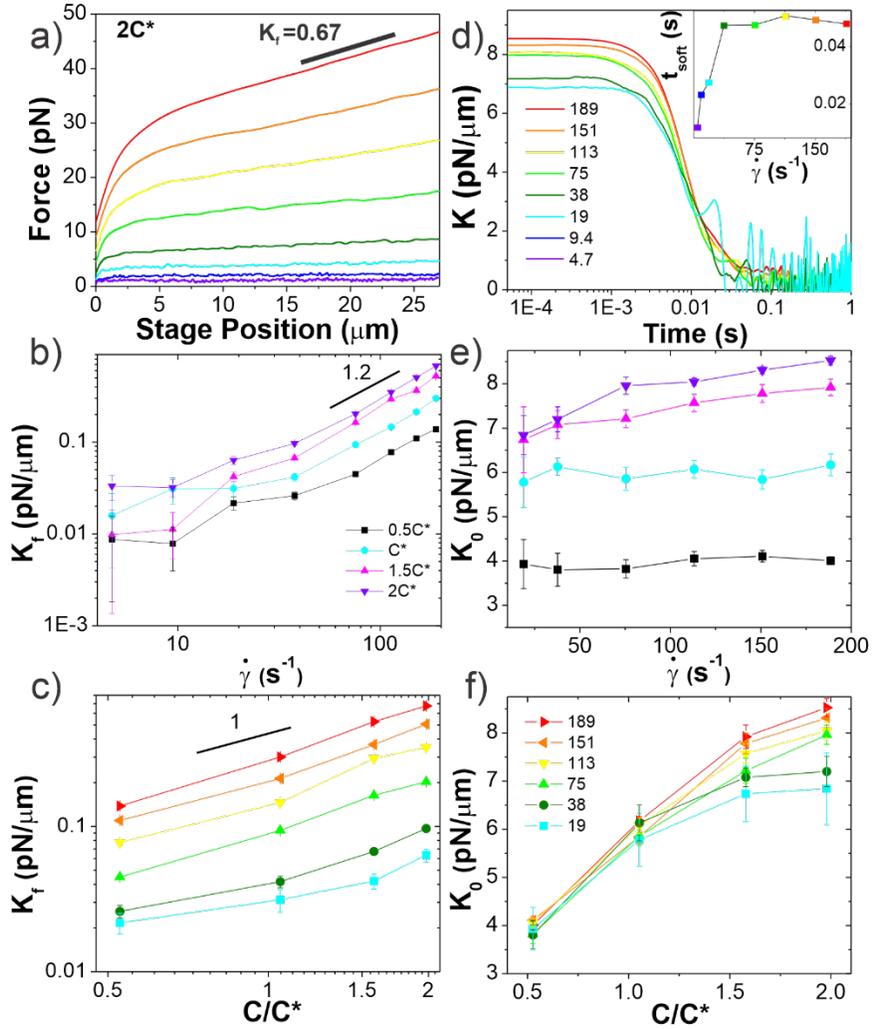

**Figure 3. Ring-supercoiled DNA blends straddling $c^*$ universally exhibit nonlinear stress response features indicative of strong entanglements.** (a) Measured force in response to strain rates of $\dot{\gamma}$=4.7–189 s$^{-1}$ as listed in (d). Data shown is for $2c^*$ (see Fig. S3 for other concentrations). (b-c) Final differential modulus $K_f$, determined from the slopes of force curves in the final response phase shown in (a) and Fig. S1, versus (b) $\dot{\gamma}$ (see legend) and (c) $c$ (see legend in (f)). (d) Differential modulus, $K=dF/dx$, as a function of time for $2c^*$ and $\dot{\gamma}$ listed in legend (other concentrations shown in Fig. S1). Inset: Average softening time $t_{soft}$, determined as the time at which $K \approx K_f$, versus $\dot{\gamma}$ for $2c^*$. (e) $K_0$ vs $\dot{\gamma}$, only showing $\dot{\gamma}$-dependence for $c>c^*$. (f) Data from (e) plotted as a function of $c$, showing a crossover from $\dot{\gamma}$-independence to $\dot{\gamma}$-dependent increase of $K_0$ at $\sim c^*$. All data in b,c,e,f have error bars but in some instances they are smaller than symbol sizes.

In order to further characterize this unexpected elasticity, we calculate an effective differential modulus $K=dF/dx$, which quantifies the stiffness of the system (Figs. 3b-f,S3). As shown in Fig. 3b,c, the $K$ value in the final response phase $K_f$, which we define as the average value in the plateau region that all $K$ curves exhibit, is nonzero and increases with $\dot{\gamma}$ and $c$ (Fig S4). The dependence of $K_f$ on $\dot{\gamma}$ demonstrates that measurements are indeed probing the nonlinear regime.[11] However, for $\dot{\gamma}<40s^{-1}$ the dependence on $\dot{\gamma}$ appears to be weaker than for higher rates which follow a power-law of ~1.2. This data suggests that lower rates may not be accessing the nonlinear regime, but may rather be in a crossover regime between linear and nonlinear response dynamics.

The initial stiffness $K_0$ displays a crossover at $c^*$ for all rates (Fig. 3d-f). Namely, $K_0$ is largely independent of $\dot{\gamma}$ for $c<c^*$ (Fig. 3f), suggestive of a linear response; while for $c \geq c^*$ $K_0$ increases with $\dot{\gamma}$, similar to the nonlinear response observed for entangled linear DNA.[11] The absence of a crossover in $K_f$ for fast rates then suggests that these large strains are sufficient to alter the interactions between polymers such that they exhibit strong entanglement-like interactions even at $c<c^*$. Nonlinear forcing has been shown to induce similar strain-induced network alterations in entangled linear polymers, due to entanglement tube dilation and contraction as well as convective constraint release.[9, 39, 43, 54-62]

To shed further light on the transition from the initial to final phase of the nonlinear response we determine the time $t_{stiff}$ at which blends deviate from the initial elastic phase (when $K$ drops to $K_0/2$), which is a measure of the fastest relaxation time of the system. We find $t_{stiff}=0.007\pm0.002$s, independent of $\dot{\gamma}$ and $c$, which agrees with the Zimm time for supercoiled constructs ($\tau_{Z,S}\approx 0.008$s). While Zimm relaxation is expected for $c<c^*$, it is rather surprising that is persists for $c>c^*$, and that there is no evidence of Zimm relaxation for rings.

We also quantify the time at which blends enter the final regime $t_{soft}$, which we define as the time at which $K$ first reaches $K_f$ (Fig S4). For all concentrations, $t_{soft}$ increases with $\dot{\gamma}$ for $\dot{\gamma}<40s^{-1}$, but for higher rates reaches a $\dot{\gamma}$-independent value of $t_{soft}\approx 0.047\pm0.004$s, quite close to the predicted Rouse time for pure ring solutions ($\tau_{R,R}\approx 0.044$s). The crossover seen at $\dot{\gamma} \approx 40s^{-1}$, similar to that observed for $K_f$, corroborates that lower rates are not well within the nonlinear regime.

The loss of substantial elasticity over $\tau_{R,R}$ indicates that Rouse-like relaxation of rings and Zimm relaxation of supercoiled molecules are the dominate modes of stress relaxation. However, because blends maintain some elasticity throughout the strain, a slower mode, such as the disengagement time $\tau_D$, must also be present. We offer one possible mechanism, which we explore further below, that could give rise to the emergent physics. Namely, because of the different relaxation timescales and conformations of rings and supercoils, nonlinear strains could force their separation, such that in the vicinity of the strain there are regions of freely diffusing supercoiled constructs that have unthreaded or untangled from rings (undergoing Zimm relaxation), and regions of pure rings that remain entangled or at least strongly overlapping (undergoing Rouse relaxation and disengagement).

Following strain, the probe is halted and the force is measured as the system relaxes (Figs. 1f,4a,S5). As with previous studies on entangled linear and ring DNA,[39, 47] a sum of up to three exponentials ($F(t)=C_1e^{-t/\tau_1}+C_2e^{-t/\tau_2}+C_3e^{-t/\tau_3}$) fits our data well (Figs. S5,S6).

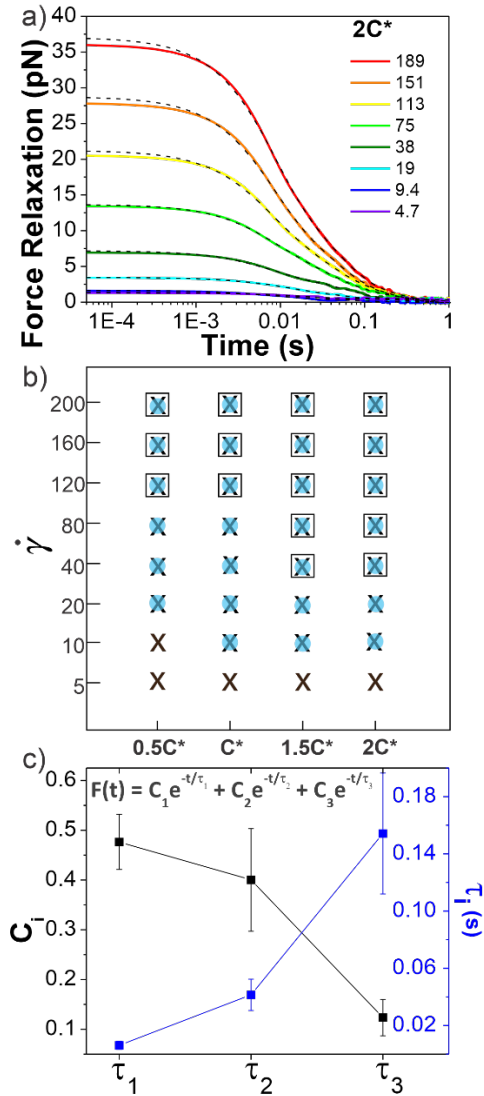

**Figure 4. Ring-supercoiled DNA blends exhibit multi-mode relaxation following nonlinear strain**. (a) Force relaxation as function of time following strain for $\dot{\gamma}$ values (s$^{-1}$) listed in legend. Each curve is fit to a sum of 1-3 exponential decays (green dashed lines, equation shown in (c)). Data shown is for $2c^*$ (other concentrations in Fig. S5). (b) Time constants from exponential decay fits for varying blend concentrations (x-axis) and $\dot{\gamma}$ (y-axis). Filled blue circles, black crosses, and open squares represent the fast ($\tau_1$), intermediate ($\tau_2$) and slow ($\tau_3$) time constants, respectively. (c) Relative coefficients $C_i$ (black) and time constants $\tau_i$ (blue), averaged over all $c$ and $\dot{\gamma}$, for each decay mode.

In all fits the different time constants are separated by close to an order of magnitude with values ~$O(10^{-3})$s, $O(10^{-2})$s, and $O(10^{-1})$s. As such, we group time constants into slow, intermediate and fast modes based on this criterion (Fig 4b), with 99% confidence intervals of $\tau_1$=0.006±0.003s, $\tau_2$=0.04±0.01s, and $\tau_3$=0.15±0.04s and corresponding relative coefficients of $C_1$= 0.48±0.05, $C_2$= 0.40±0.10 and $C_3$= 0.12±0.04 (Fig 4c). For $\dot{\gamma}$<40s$^{-1}$, single or double exponentials with time constants of $\tau_1$ (single) or $\tau_1$ and $\tau_2$ (double) are sufficient to describe the data, suggesting that the slowest relaxation mode is distinct to the nonlinear regime.

$\tau_1$ is nearly identical to $t_{stiff}$ measured during strain, corroborating that Zimm relaxation of supercoils is the fastest nonlinear relaxation mode. To understand the relaxation mechanisms associated with the two slower modes we compare our measured values to the predicted and measured values of $\tau_R$ and $\tau_D$ for linear and ring polymer systems. We find that $\tau_2$ and $\tau_3$ are comparable to $\tau_R$ and $\tau_D$ for ring DNA ($\tau_{R,R}\approx0.04$s, $\tau_{D,R}\approx0.1$s), but significantly shorter than those for linear DNA ($\tau_{R,L}\approx0.13$s, $\tau_{D,L}\approx1.24$s).[11, 41, 46, 47] By comparing the contributions from each mode, we see that the system relaxes mainly through apparent Zimm relaxation of supercoils with $C_1\approx48\%$ and Rouse-like relaxation with $C_2\approx40\%$ (Fig. 4b,c). This result is in line with our $t_{soft}$ analysis that shows that blends dissipate most of their elastic stress on the order of $\tau_{R,R}$ despite the existence of a slower relaxation mode and sustained elastic response to strain.

It is worth discussing the differences between our nonlinear and linear microrheology data. As described above, we attribute the agreement of linear microrheology results with those of entangled linear polymers to interactions between the two topologies that cause substantial entanglements – similar to comparable linear polymer systems – even in semidilute conditions. However, the agreement of $t_{soft}$ and $\tau_2$ with $\tau_{R,R}$ rather than $\tau_{R,L}$, and likewise $\tau_3$ with $\tau_{D,R}$ rather than $\tau_{D,L}$, suggest that nonlinear forcing is sufficient to alter the interactions between topologically distinct polymers such that they lose blend characteristics and behave closer to pure entangled ring polymer solutions. At the same time, while the existence of multiple relaxation modes is in line with our linear regime results that show that for $c>c^*$ blends behave as if entangled, in the nonlinear regime these modes persist for all concentrations. These results support our suggested mechanism of nonlinear straining separating supercoils from rings and forming separate regions of densely entangled rings and minimally overlapping supercoils. Similar behavior has been seen for entangled linear DNA in which nonlinear micro-strains compress polymers in front of the moving bead, thereby increasing the local entanglement density while leaving dilute regions in its wake.[63] This effect may also explain the emergent sustained elasticity.

**CONCLUSIONS**

In conclusion, we present linear and nonlinear rheological properties of blends of relaxed circular and supercoiled DNA at concentrations that straddle the overlap concentration. Surprisingly, despite being in the dilute-semidilute crossover regime, we observe dynamics indicative of entanglements, which we suggest arise from synergistic interactions between the two topologies. Linear microrheology reveals a crossover at $c^*$ from semidilute dynamics to those that align with entangled linear polymers. At the same time, nonlinear microrheology uncovers unique sustained elasticity and multiple relaxation modes not expected at these modest concentrations. Interestingly, while blends exhibit linear viscoelasticity comparable to those of entangled linear polymers, nonlinear response characteristics align more closely with predictions for entangled rings. We interpret these differences as arising from strain-induced network rearrangements that alter the entanglement density and disrupt the interactions between topologically-distinct polymers.

In summary, our results reveal that blended solutions of ring and supercoiled polymers exhibit unexpected viscoelastic properties at surprisingly low concentrations. As a result, this study is not only of fundamental importance to polymer physics research but also has commercial applications. Namely, topological blends can potentially be exploited as a route for designing low-mass high-strength viscoelastic materials. Finally, we hope the new phenomena we report spur theoretical investigations into similar topological blends to shed light onto the physical interactions between topologically distinct polymers that give rise to the emergent dynamics they exhibit.

## MATERIALS AND METHODS

Complete experimental details, summarized below, are provided in SI.

Circular 50 kbp DNA was prepared using protocols detailed elsewhere.[5, 64] The purified solution has a concentration of 0.56 mg/mL and consisted of ~69% relaxed circular, ~26% supercoiled, and ~5% linear DNA, as quantified via single-molecule 'counting' experiments (SI, Fig S1). We determine $c^*\approx 0.26$ mg/mL using a weighted average of $R_G$ values for relaxed circular, supercoiled and linear species.[4, 5] While blends contain a small fraction of linear chains, based on our previous work we do not expect these contaminants to significantly impact our results (see SI for details), so we treat our blend as a two-component system comprised of rings and supercoiled constructs.

Microrheology measurements are described in Figure 1 and SI.

## FOOTNOTES

**Supporting Information.** Expanded experimental section; Figure S1. Single-molecule 'counting' experiments to visualize DNA molecules in the blend; Figure S2. Position and velocity of nanopositioning stage during active microrheology measurement; Figure S3. Nonlinear force response of blends of ring and supercoiled DNA; Figure S4. Final phase of differential modulus in response to nonlinear strains; Figure S5. Relaxation of force induced in ring-supercoiled DNA blends following nonlinear strains. Figure S6: Initial force relaxation following nonlinear strain.

## AUTHOR INFORMATION

**Corresponding Author**

*Email: randerson@sandiego.edu

## ORCID

Rae M. Robertson-Anderson: 0000-0003-4475-4667

## CONFLICTS OF INTEREST

There are no conflicts to declare.

## ACKNOWLEDGEMENTS

The authors acknowledge financial support from Air Force Office of Scientific Research (AFOSR-FA9550-17-1-0249) and National Science Foundation (NSF-CBET-1603925).

**Table of Contents artwork**

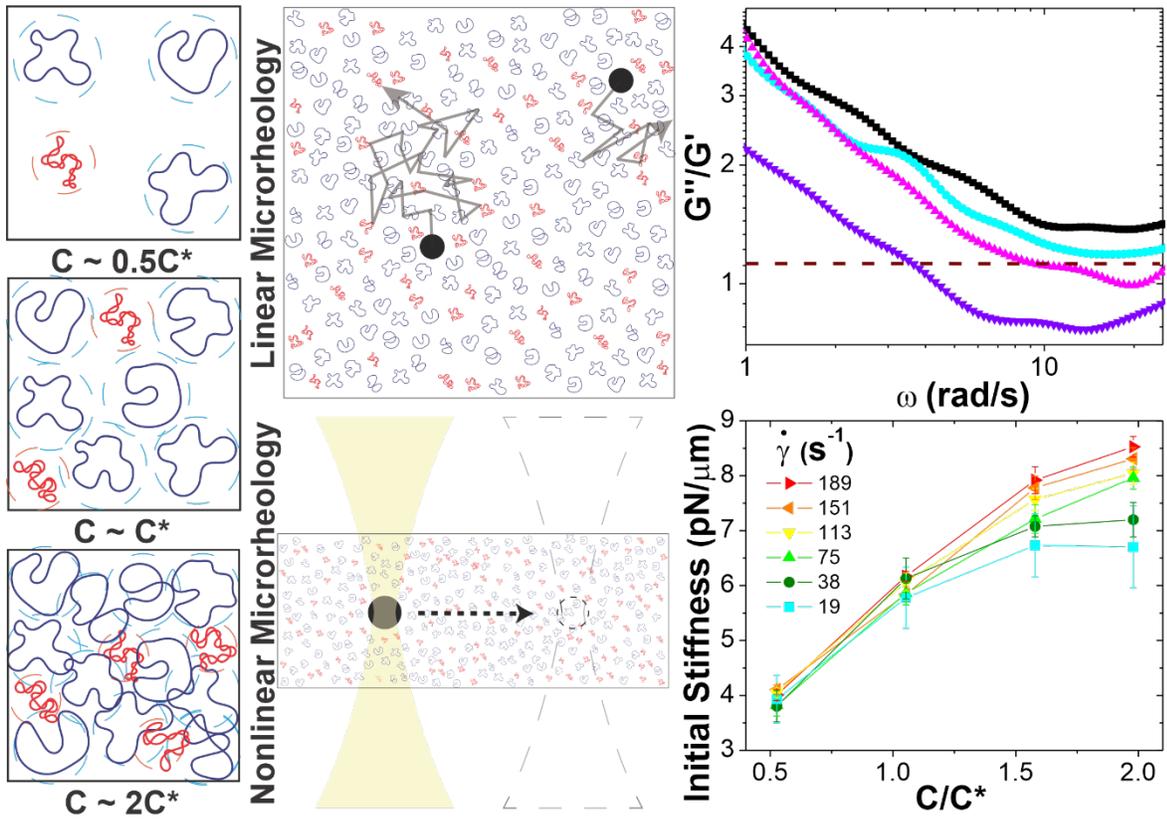

**Unexpected entanglement dynamics in semidilute blends of supercoiled and ring DNA**

Karthik R. Peddireddy[1], Megan Lee[1], Yuecheng Zhou[2], Serenity Adalbert[1], Sylas Anderson[1], Charles M. Schroeder[2], Rae M. Robertson-Anderson[1,*]

[1]*Department of Physics and Biophysics, University of San Diego, 5998 Alcala Park, San Diego, CA 92110, United States*

[2]*Department of Materials Science and Engineering, Beckman Institute for Advanced Science and Technology & Department of Chemical and Biomolecular Engineering, University of Illinois at Urbana-Champaign, Urbana, IL 61801, United States*

**Supporting Information**

**Expanded Experimental Section**

**DNA Preparation:** Circular 50 kbp DNA molecules were prepared by replication of fosmid constructs in Escherichia coli, followed by extraction, purification and enzymatic treatment using protocols detailed elsewhere[1,2] and briefly described below.

To replicate DNA, E. coli cultures containing the fosmid clone were grown from frozen glycerol stocks. To extract the DNA, cells were lysed via treatment with an alkaline solution. The extracted DNA was then renatured via treatment with an acidic detergent, precipitated in isopropanol, washed with 70% ethanol, and resuspended in TE buffer (10 mM Tris-HCl (pH 8), 1 mM EDTA).

To purify the DNA, the solution was treated with Rnase A (to remove contaminating RNA) followed by phenol-chloroform extraction and dialysis (to remove proteins). The DNA was stored in TE10 buffer (10 mM Tris-HCl (pH 8), 1 mM EDTA, and 10 mM NaCl) at 4°C.

**DNA Blend Characterization:** The resulting 50 kbp DNA solution was analyzed via agarose gel electrophoresis to determine solution concentration and estimate the percentage of different DNA topologies comprising the sample. Gel analysis was performed using a Life Technologies E-Gel Imager and Gel Quant Express software. From the gel analysis, we determined that the purified solution had a total DNA concentration of 0.56 mg/mL and consisted of ~75% relaxed circular (*R*) molecules and ~25% supercoiled (*S*). We note that gel electrophoresis cannot accurately detect small (<5%) fractions of molecules in a sample, and loading the sample into the microscope chamber for experiments can introduce shear-induced nicks and cuts in the DNA that may not be present in the gel results.

To improve the accuracy of our blend composition determination, and determine the structure and heterogeneity of supercoiled constructs, we performed single-molecule 'counting' experiments described below. In these experiments, we directly imaged 231 randomly chosen molecules in the sample under flow to determine the topology of each molecule and establish sufficient statistics to determine the overall fraction of ring, supercoiled, and linear molecules in the sample. This method also allowed us to determine the structure that supercoiled constructs assumed (i.e. linear, branched, etc).

For counting experiments, 5 μL of the DNA blend was diluted 50x and fluorescently labeled with YOYO-1 (ThermoFisher) at a dye-to-base pair ratio of 1:4. A trace amount of labeled molecules was added to a viscous imaging buffer solution containing 10 mM Tris-HCl (pH 8), 1 mM EDTA, 10 mM NaCl and 60% w/w sucrose. In addition, a small amount of reducing agent β-mercaptoethanol (14 μM) and coupled enzymatic oxygen scavenging system containing glucose (50 µg/mL), glucose oxidase (0.01 µg/mL), and catalase (0.004 µg/mL) were added to suppress photobleaching and photocleaving. The sample was then rotationally mixed for >20 minutes before introducing into the microfluidic cross-slot device, fabricated using standard techniques in soft lithography as described before.[3] In brief, the microfluidic device contained a fluidic layer situated below a control layer containing a fluidic valve. The fluidic layer was fabricated to contain a cross-slot channel geometry to generate planar extensional flow. All DNA molecules were then stretched under the same flow strength. Single-molecule imaging was performed using an inverted epifluorescence microscope (IX71, Olympus) and EMCCD camera (iXon, Andor Technology). Labeled DNA solutions were illuminated using a 50 mW 488 nm laser (Spectra-Physics, CA, USA) directed through a 2.2 neutral density filter (ThorLabs, NJ, USA) and a 488 nm single-edge dichroic (ZT488rdc, Chroma). Fluorescence emission was collected by a 1.45 NA, 100× oil immersion objective lens (UPlanSApo, Olympus) and 1.6× tube lens giving a total magnification of 160×. A 525 nm bandpass filter (FF03-525/50-25, Semrock) was used in the detection path. Finally, images (512×512 pixels, 16 µm pixel size) were acquired under frame transfer mode at 33 fps.

In general, for relaxed rings the two strands can be discerned, especially during the relaxation phase after the cessation of planar extensional flow (Fig. S1, top); while for linear DNA molecules, the contour length is twice that of their ring counterparts making them easily distinguishable from ring DNAs (Fig. S1, bottom). Supercoiled molecules all exhibit single 'ribbon-like' structures, which are essentially linear rather than branched, after being stretched (Fig. S1, middle). By comparing the extended conformations of supercoiled and linear molecules of the same length $L$ we can see that the supercoiled DNA 'length' appears to be comparable to the $0.4L$ value reported in Ref 4 that we use to compute $R_G$ and thus $c^*$ (described in following section).

Beyond revealing the structures assumed by the topologically-distinct molecules, the results of our single-molecule analysis yielded a blend composition of ~69% rings, ~26% supercoiled molecules, and 5% linear chains. While this makeup is close to that obtained via gel electrophoresis it does reveal the presence of a small fraction of linear chains. However, based on our previous steady-state diffusion studies for ring-linear DNA blends[5], we do not expect that this small fraction plays a significant role in the results we present. In the referenced study (Ref 5) we showed that for a comparable DNA length and concentration (45 kbp, 0.5 mg/ml), the introduction of 5% linear chains into a ring DNA solution only reduced the diffusion of ring and linear DNA by ~3% and ~12% respectively. This is compared to the 21% and 51% drop measured at ~25%. Because the supercoiled contaminants, which appear to assume conformations akin to shorter linear chains, make up >25% of the blend, we conclude that it is the presence of supercoiled constructs rather than linear chains that play the dominant role in the intriguing mechanics we report.

Further, we note that the pipetting method to introduce DNA molecules into the flow device, as well as the extra handling involved in labeling molecules, can introduce nicks in rings, thus overestimating the fraction of linear chains. In microrheology experiments, we use wide-bore pipet tips (not possible for microfluidics experiments) and only require a single pipetting step. As such, the 5% linear chains that single-molecule experiments measure is quite likely an overestimate. We nonetheless take it into account when computing the overlap concentration (see below) and interpreting our experimental results.

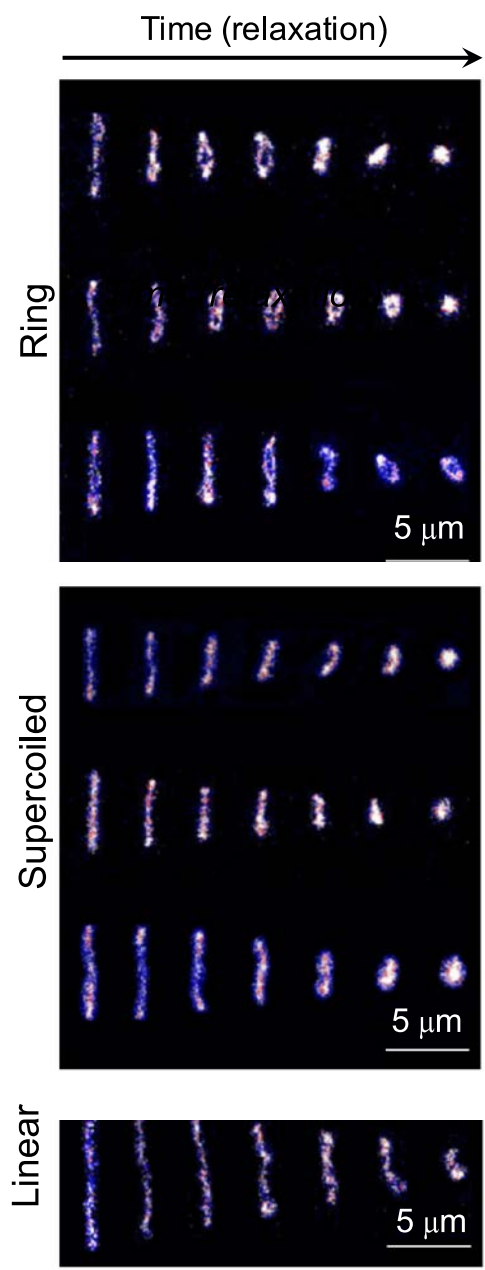

**Figure S1. Single-molecule 'counting' experiments to visualize DNA molecules in the blend.** Single-molecule snapshots of 50 kbp relaxed ring (top), supercoiled (middle) and linear (bottom) DNA molecules within the same sample during the relaxation phase after stretching under planer extensional flow. 231 molecules in total were stretched and imaged during relaxation to determine that ring, supercoiled and linear molecules comprise 69%, 26% and 5% of the blend.

**Overlap Concentration Determination:** Measurements were performed with dilutions of this sample to 0.14, 0.27, 0.41 and 0.51 mg/mL, chosen to span from ~2x below to ~2x above $c^*$ (Fig. 1a). We compute $c^*$ for blends using a weighted average of $R_G$ values for relaxed circular, supercoiled, and linear species determined as described below.

The radius of gyration for rings has been shown to be smaller than their linear counterparts with a ratio $R_{G,L}/R_{G,R}$ =1.58 measured for DNA.[1] The radius of gyration for supercoiled DNA ($R_{G,S}$) has likewise been shown to be smaller than linear chains and can be calculated via the worm-like-chain expression for linear polymers assuming a contour length of $L_S$=0.4$L$, where $L$ is the contour length of the polymer:

$$R_{G,S} = p\left[\frac{0.4L}{3p} - 1 + 2\left(\frac{p}{0.4L}\right)^2 \left(1 - e^{-0.4L/p}\right)\right]^{0.5},$$

where $p$ is the persistence length (~50 nm for DNA).[4] We note that the 0.4 prefactor may depend on ionic strength (the ionic strength used in Ref 4 is higher than our conditions), and may overestimate $R_G$ for larger DNA (as noted by authors of Ref 4). However, by comparing the extended conformations of supercoiled and linear molecules of the same length $L$ in our blends (Fig S1) we can see that the supercoiled DNA 'length' appears to be comparable to the 0.4$L$ value reported in Ref 4 that we use to compute $R_G$ and thus $c^*$.

Using the expressions and relations for $R_G$ for both topologies provided above, along with reported values of $R_G$ for similarly sized ring and linear DNA, we compute $R_{G,S} \cong 0.33$ μm and $R_{G,R} \cong 0.37$ μm ($R_{G,L} \cong 0.56$ μm). From these values we calculate an effective $R_{G,blend}$ of 0.37 μm and corresponding $c^*$ value of ~0.26 mg/mL. As such, our chosen concentrations equate to ~$0.5c^*$, $c^*$, $1.5c^*$ and $2c^*$. For reference, $c^*$ for the equivalent system of linear DNA is ~4x smaller (~74 μg/ml). Throughout the manuscript we refer to all blend concentrations in terms of $c^*$.

Using $R_{G,blend}$ we can also compute the mesh size $\xi$ of the different blends via $\xi = \sqrt{6}R_G(c/c^*)^{-3/4}$.[6] The mesh sizes for the 4 different blends are 1.44 μm (0.14 mg/ml), 0.88 μm (0.27 mg/ml), 0.64 μm (0.41 mg/ml), and 0.55 μm (0.51 mg/ml).

**Sample Preparation:** For passive and active microrheology, 1 μm and 4.5 μm carboxylated polystyrene microspheres (Polysciences, Inc.) were added to solutions, respectively. Both beads were coated with Alexa-488 BSA to prevent DNA adsorption and enable fluorescence visualization. To inhibit photobleaching of microspheres, glucose (45 μg/mL), glucose oxidase (43 μg/mL), catalase (7 μg/mL) and β-mercaptoethanol (5 μg/mL) were added. 0.1% Tween-20 was also added to prevent DNA adsorption to sample chamber surfaces. Using a wide-bore pipet tip to avoid shearing, the resulting solution was pipetted into a sample chamber comprised of a microscope slide and coverslip with two pieces of double-stick tape in between. The chamber was then sealed with epoxy and allowed to equilibrate for ~15 mins before measurements.

**Passive Microrheology:** For passive microrheology measurements (Fig. 1b-d), diffusing microspheres were visualized using an Olympus IX73 microscope with a 20x objective and high-speed CMOS camera (Hamamatsu Orca Flash 2.8). For each concentration, 15 time-series of 512x512 (181 nm/pixel) images consisting of ~150 beads per frame were recorded for 15 seconds at 30 fps. Custom-written MATLAB code was used to extract the trajectories of diffusing beads and calculate the mean-squared displacements (MSD) in the $x$ and $y$ directions. All MSDs shown consist of ~2000 particles and are an average of MSDs in $x$ and

$y$ directions, denoted as $<\Delta r^2(t)>$. Diffusion coefficients were calculated via $<\Delta r^2(t)>=2Dt$; and linear viscoelastic moduli ($G'(\omega)$, $G''(\omega)$) were determined via the generalized Stokes-Einstein relation:[7]

$$G^*(\omega) = G'(\omega) + iG''(\omega) = \frac{k_B T}{i\omega <\Delta r^2(\omega)> \pi R},$$

where $k_B$ is Boltzmann's constant, $T$ is the absolute temperature, $<\Delta r^2(\omega)>$ is the Fourier transform of $<\Delta r^2(t)>$, and $R$ the radius of the beads. The Fourier transform of $<\Delta r^2(t)>$ is obtained by:[8]

$$-\omega^2 <\Delta r^2(\omega)> = (1-e^{-i\omega t_1})\frac{<\Delta r^2(t_1)>}{t_1} + 2De^{-i\omega t_N} + \sum_{k=2}^{N}\left(\frac{<\Delta r^2(t_k)>-<\Delta r^2(t_{k-1})>}{t_k-t_{k-1}}\right)(e^{-i\omega t_{k-1}} - e^{-i\omega t_k}),$$

where 1 and $N$ in the equation represent the first and last point from the oversampled MSD data. Oversampling is done using the PCHIP MATLAB function. More details about the data analysis can be found in Ref [9].

While we acquire particle trajectories over the time interval [0.033 – 15 s] corresponding to [0.42 – 188 rad/s], we only use the interval [0.25 – 15 s] in our analysis as for shorter times the frame-to-frame bead displacements are not significantly larger than the precision in centroid localization (~200 nm vs 45 nm).[10] At 0.25 s, the displacements are an order of magnitude larger than the tracking precision of ~45 nm.

**Active nonlinear microrheology:** We use optical tweezers to apply fast mesoscale strains to the blends (Fig. 1e,f). The optical trap consists of an Olympus IX70 microscope with a 60$x$ 1.4 NA objective (Olympus) and a 1064 nm Nd:YAG fiber laser (Manlight). A position sensing detector (Pacific Silicon Sensors) is used to measure the deflection of the laser beam, which is proportional to the force exerted on the trapped bead. The proportionality constant (i.e. trap stiffness) is obtained via Stokes drag method as previously described.[11, 12] Strains are applied to blends by moving a nanopositioning piezoelectric microscope stage (Mad City Labs) a fixed distance of 30 μm (Fig. 1e) at speeds of $v = 5 – 200$ μm/s, which are converted to strain rates via $\dot{\gamma}=3v/\sqrt{2}R$.[13] Both stage position and laser deflection data are acquired at 20 kHz before (5 s, equilibrium), during (0.15-6 s, strain) and following (9-15 s, relaxation) the strain (Fig. 1f).

The nanopositioning stage takes 0.002 s to accelerate to constant speed from rest and decelerate to rest following constant rate strain (Fig. S2). As such, the force data we show during the strain phase is only for the portion of the strain that is at constant speed (chopping off the initial and final 0.002 s of data). Likewise, the relaxation data shown starts 0.002 s after the stage begins to stop (once it has come to complete halt).

At least 15 trials, each with a new bead in a new unperturbed location, are conducted for every speed and concentration in order to verify homogeneity throughout the sample and reproducibility of the force values. All displayed data is the average and standard error of all trials for each condition.

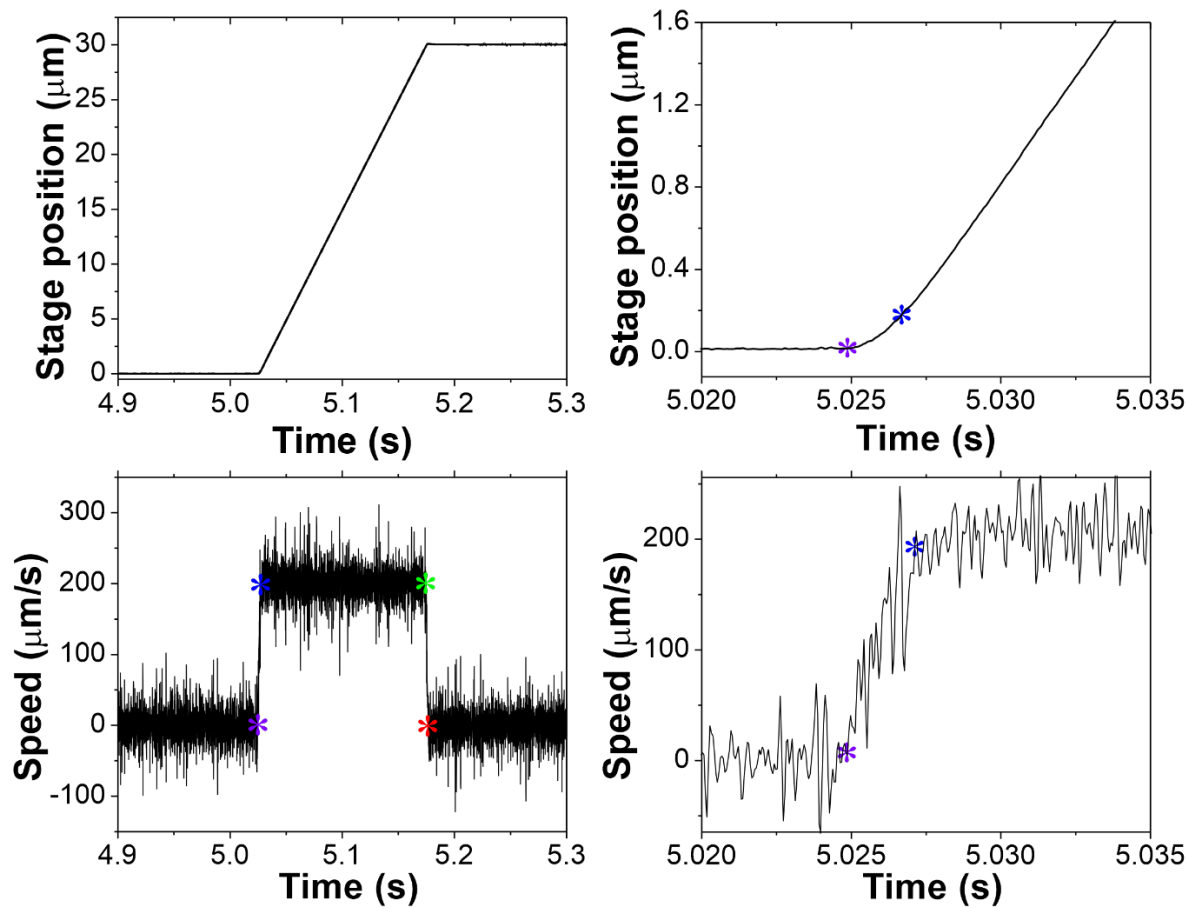

**Figure S2. Position and velocity of nanopositioning stage during active microrheology measurement.** (Top left) Stage position during all phases of a 200 μm/s strain experiment. (Top right) Zoom-in of the start of the stage motion. A constant slope (i.e. 200 μm/s) is achieved in ≤0.002 s (indicated by violet and blue stars). (Bottom left) Stage velocity during all phases of a 200 μm/s strain experiment. For the strain phase, we evaluate force curves during the time at which stage maintains constant speed (indicated by blue and green stars. For the relaxation phase, we evaluate force curves once the stage comes to a complete stop (indicated by red star). (Bottom right) Zoom-in of the start of the strain phase. As shown, the stage takes ≤0.002 s to reach constant speed from rest. Stage takes identical time to come to rest following constant speed strain.

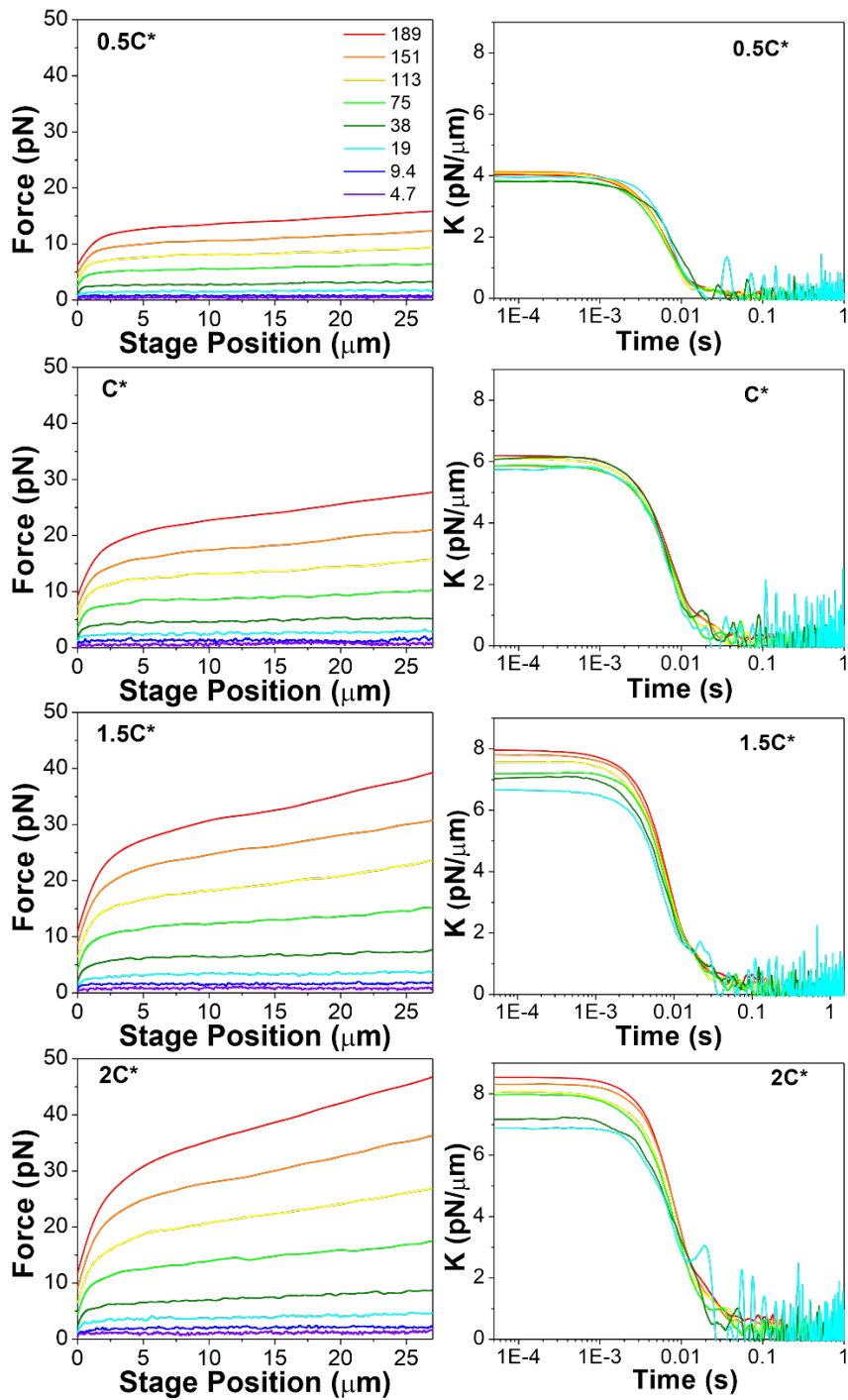

**Figure S3. Nonlinear force response of blends of ring and supercoiled DNA.** Measured force (left) and corresponding differential modulus, $K=dF/dx$, (right) in response to 30 μm strains with $\dot{\gamma}$ listed in units of s$^{-1}$ in legend (top left). The blend concentration is listed in units of $c^*$ at the top of each plot. For a given concentration, force increases with increasing $\dot{\gamma}$. Similarly, for a given $\dot{\gamma}$, force increases with increasing $c$. For $c<c^*$, dependence of $K$ on $\dot{\gamma}$ is negligible but becomes significant for $c>c^*$. At any given $\dot{\gamma}$, initial $K$ value increases with increasing concentration.

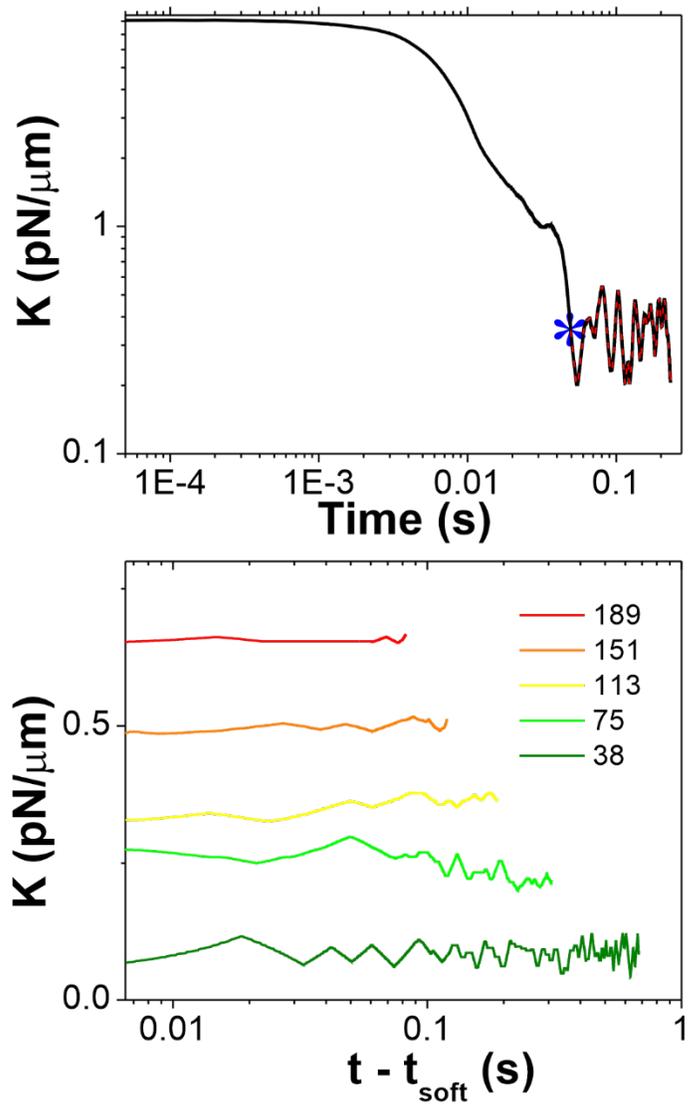

**Figure S4. Final phase of differential modulus in response to nonlinear strains.** (Top) A sample $K$ vs $t$ curve ($2c^*$, $\dot{\gamma}=113$ s$^{-1}$). Blue star indicates the time we define as $t_{soft}$ and red dashes indicate the region averaged over to compute $K_f$. (Bottom) The final plateau regions of $K$ vs $t$ curves for $2c^*$, smoothed using a 2500-5000 point moving median (MATLAB). $\dot{\gamma}$ are listed in units of s$^{-1}$ in legend.

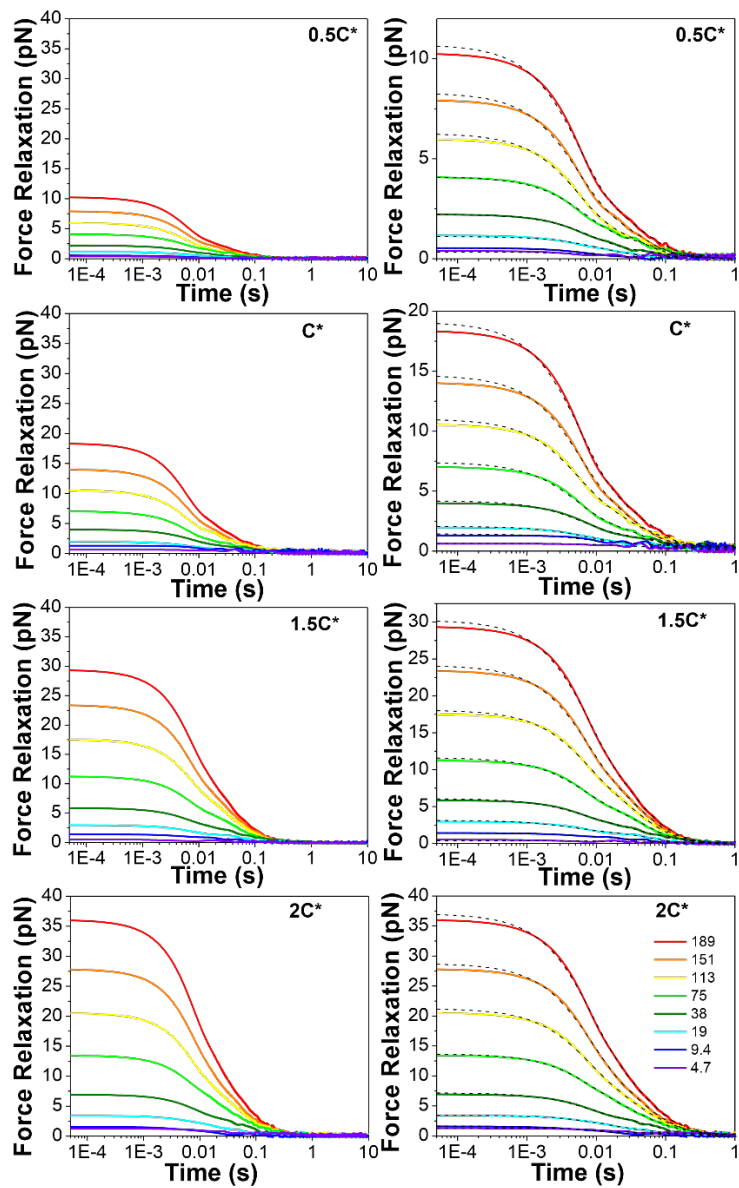

**Figure S5. Relaxation of force induced in ring-supercoiled DNA blends following nonlinear strains.** (Left) Measured force relaxations following 30 μm strains with $\dot{\gamma}$ listed in units of s$^{-1}$ in legend (top plot). (Right) Zoom-ins of relaxations with corresponding fits to exponential decay functions (black dashed lines). The blend concentration is listed in units of $c^*$ at the top of each plot. For $\dot{\gamma}>40$ s$^{-1}$, relaxations are well-fit to a sum of three exponential decays (i.e. $F(t)=C_1e^{-t/\tau_1}+C_2e^{-t/\tau_2}+C_3e^{-t/\tau_3}$). For lower strain rates, single and double mode exponential decay functions are sufficient to fit the data (see Fig S6).

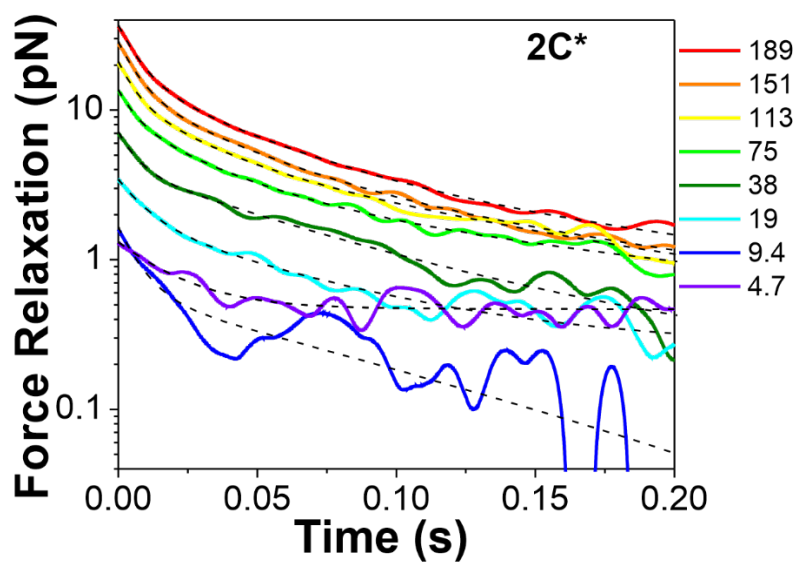

**Figure S6. Initial force relaxation following nonlinear strain.** Bottom-right plot of Fig S5 replotted on linear-*x* log-*y* scale and truncated to 0.20 s. Dashed lines are corresponding fits to exponential decay functions. $\dot{\gamma}$ listed in units of s$^{-1}$ in legend (top plot). Note that force decay begins immediately and is initially fast ($\tau_1$ and $\tau_2$ relaxation) then slows to a steady exponential decay (constant slope, $\tau_3$).